\documentclass[letterpaper]{article}
\usepackage{aaai16}
\usepackage{times}
\usepackage{helvet}
\usepackage{courier}
\usepackage{balance}  
\usepackage{graphics} 
\usepackage{txfonts}
\usepackage{times}    
\usepackage{url}      
\usepackage{color}
\usepackage{textcomp}
\usepackage{booktabs}
\usepackage{ccicons}
\usepackage{todonotes}

\usepackage{epigraph}

\def\pprw{8.5in}
\def\pprh{11in}

\begin{document}

\title{Sequential Voting Promotes Collective Discovery \\ in Social Recommendation Systems}

\author{L. Elisa Celis$^\dagger$, Peter M. Krafft$^\ddagger$, Nathan Kobe$^\dagger$
 \\ $^\dagger$\'Ecole Polytechnique F\'ed\'ereal de Lausanne, Switzerland
 \\ $^\ddagger$Massachusetts Institute of Technology, USA
\\ elisa.celis@epfl.ch, pkrafft@mit.edu, nathan.kobe@epfl.ch
   }

\maketitle

\begin{abstract}

One goal of online social recommendation systems is to harness the wisdom of crowds in order to identify high quality content. Yet the sequential voting mechanisms that are commonly used by these systems are at odds with existing theoretical and empirical literature on optimal aggregation. This literature suggests that sequential voting will promote herding---the tendency for individuals to copy the decisions of others around them---and hence lead to suboptimal content recommendation. Is there a problem with our practice, or a problem with our theory? Previous attempts at answering this question have been limited by a lack of objective measurements of content quality. Quality is typically defined endogenously as the popularity of content in absence of social influence.  The flaw of this metric is its presupposition that the preferences of the crowd are aligned with underlying quality. Domains in which content quality can be defined exogenously and measured objectively are thus needed in order to better assess the design choices of social recommendation systems. In this work, we look to the domain of education, where content quality can be measured via how well students are able to learn from the material presented to them. Through a behavioral experiment involving a simulated massive open online course (MOOC) run on Amazon Mechanical Turk, we show that sequential voting systems can surface better content than systems that elicit independent votes.

\end{abstract}

\section{Introduction}

Many social media sites, news aggregators, Q\&A websites, massive open
online courses (MOOCs), and other online platforms such as Amazon,
Yelp and YouTube rely on social recommendation in order to aggregate the
opinions of their many users.  Content is then sorted or ranked
using this information.  Recommendations are solicited by, e.g.,
allowing users to like, rate, or, when presented with
items in a ranked order, upvote content, and  aggregate
opinions are often displayed publicly. Such sequential voting systems, in
which individual or aggregated previous votes are publicly displayed and
continuously updated, are now widespread on the internet.

Yet literature on collective intelligence and the wisdom of crowds
suggests that these sequential voting systems may be suboptimal at
surfacing high quality content.  Sequential voting produces an avenue
through which previous decisions can influence future decisions.
Social influence through this avenue has been shown to exist across a
diversity of online systems, including social recommendation
systems~\cite{salganik2006experimental,muchnik2013social,van2014field}.
However, the classic Condorcet jury theorem and related results 
suggest that independent votes will yield better aggregation than
correlated ones \cite{ladha1995information}. Further theoretical
work on ``herding'' and ``information cascades'' also suggests that
sequential votes can lead to group convergence on suboptimal outcomes
\cite{bikhchandani1992theory,chamley2004rational}.  Empirical work on
aggregation for point estimation tasks has also lent 
support to these theoretical results \cite{lorenz2011social}.  In
summary, online ratings can ultimately be very different when votes
are sequential instead of independent, and the prevailing hypothesis
is that these differences may be detrimental in the sequential setting.

However, recent work has begun to question the universality of this
prediction. Researchers have investigated whether social influence
actually leads to miscalibrated rankings \cite{stoddard2015popularity}
as well as whether social influence can sometimes help improve
rankings \cite{lerman2014leveraging,abeliuk2015benefits}.  The results
thus far support the idea that sequential votes may not do much harm
to collective outcomes in social recommendation systems and leave open 
the possibility that mechanisms that allow social influence may in fact accelerate the discovery of good content and allow for effective rankings.

In order to assess the outcomes of sequential voting systems, this
line of work makes one simple but potentially problematic assumption.
They assume that content quality is equivalent to popularity as
measured by independent private votes. However, there is no reason a
priori to privilege popularity in the independent context over the
sequential one, and doing so precludes the possibility that social
interaction might ultimately surface qualitatively better
content. Domains in which content quality can be defined exogenously
and measured objectively are thus needed in order to better assess the
design choices of social recommendation systems.

\subsection{Our Contribution}

To address this problem, we select a domain in which we have an
objective measure of underlying quality: the education domain. Here,
quality is measured by how well students are able to learn from
the material presented to them. We measure learning in two ways: by
testing students on material related to the content they consume, and
by asking students to self-report their skill level before and after
consuming content.\footnote{Self-reported learning can be a proxy for
  motivation \cite{S2003}, and is correlated with actual learning
  \cite{A1999}.} This gave us two exogenous metrics of quality, which
allow us to objectively compare  different content ranking methods.

We then designed an experiment to test the effect of sequential
  voting on the quality of surfaced content.  We compare against
independent voting, in which participants cannot view previous votes.  We further benchmark against content curated by experts
and randomly selected content. Based on the convergence of the web industry on the use of sequential voting systems, we posit that sequential
voting can actually lead to better content ranking than
independent voting.

We find that self-reported learning of content consumers was
significantly higher when the content was chosen by sequential votes
compared to all other conditions, including the expert condition.  {Consistent with this trend, average
test scores are also always higher in the sequential condition, though
not statistically so.}  
Hence, sequential voting can allow for the discovery of high-quality content without expert intervention, suggesting a beneficial effect of social influence.

\section{Approach}

Our goal is to determine if sequential voting can select better answers in a student forum than other types of curation methods, including independent votes and expert-grading. The type of student forum we consider is common in MOOCs and other online Q\&A sites.  It begins with a single question (often posted by a user) followed by a sequence of answers posted by peers.  These answers are then voted on by other students. {Rather than using an existing live MOOC, we take an approach that has been used recently in the education literature to better isolate aspects of learning, e.g., \cite{CLFHH2015} and use a simulated MOOC using workers from Amazon Mechanical Turk (AMT)} as students. In this manner we {simulated} a  MOOC forum in which content was generated, ranked and consumed by AMT workers as participants.

Our goal was to quantitatively measure each participant's {skill} after
being exposed to filtered peer-generated content. This was measured by
1) an exam on the material and 2) by two survey questions answered at the end of the entire course that asked participants to self-report their knowledge before and after course completion. Each course progressed through three phases: 
\begin{itemize}
\item Content generation -- which simulates students answering questions in a forum,
\item Vote  collection -- which simulates the discovery of content from the generation phase via either sequential or independent upvotes, and 
\item Testing  -- which allows us to measure students'
knowledge after consuming content surfaced in the voting 
phase.
\end{itemize} 
See the Experiment section for more details.

Importantly, as is the case in many social recommendation systems, having a content generation phase allowed us to present material generated by peer participants. 
We were then able to study the effect that sequential vs independent voting in the collection phase had on the testing phase. Participants were allowed to engage in at most one of the three phases in each course in order to isolate the learning that occurs within each phase. 

We hypothesized that students shown content discovered by sequential votes would outperform those shown content discovered by independent votes due to an underlying herding effect that allows for greater efficiency in identifying good explanations.\footnote{Formal models for describing and predicting herd behavior have emerged in many fields (see \cite{RCF2009} for a summary); the exact mechanism through which herding or social influence occurs is immaterial to our study, we simply allow for it (or not) by showing (or hiding) votes.} In addition, we compared against benchmarks consisting of a random condition, in which random answers were shown, and an expert  condition, in which the authors graded each explanation and explanations with top grades were shown.

\section{Experiment}

We ran two simulated MOOC courses; one on introductory programming, and another on introductory art styles. Each course had five lessons: the Computer Science course taught Assignment, If-Else Statements, For Loops, Functions and Recursion, while the Art course taught the defining features of Photorealism, Realism, Expressionism, Impressionism, Abstract, Cubism, Color-Field, Hard-Edge, Nanga and Mughal. 
The authors are experts in computer science, and one of the authors is knowledgeable in art. The lessons and questions for the Computer Science course were developed jointly, and the lessons and questions for the Art course were developed by the knowledgeable author. In order to validate the authors' decisions, two professors in Computer Science and one professor in Art, each with more than 15 years experience teaching introductory courses in their respective areas, were consulted, as described in the External Validators section.

\begin{figure}[t!]
\centering
  \includegraphics[width=.8\columnwidth]{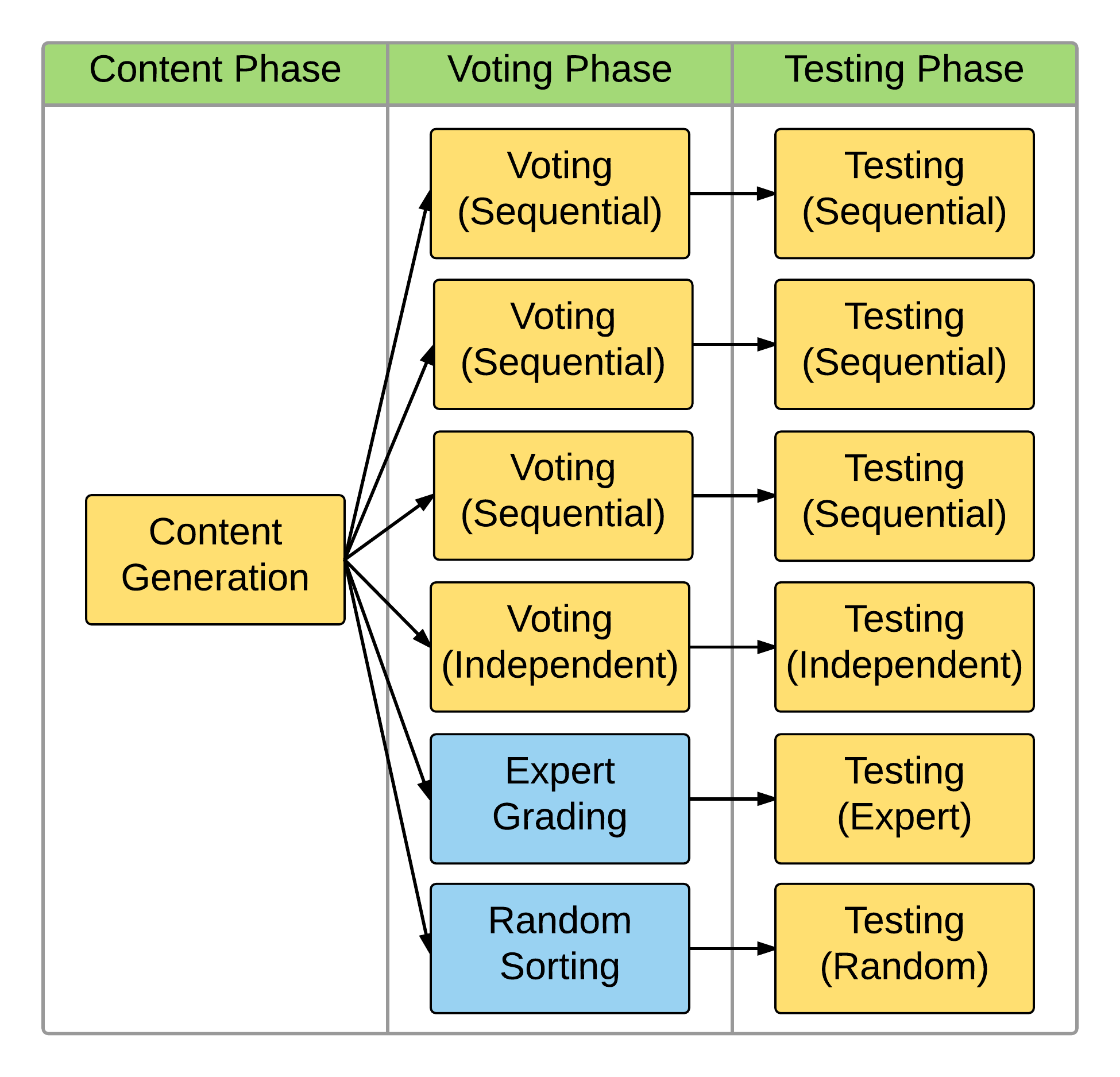}
  \caption{The experimental design for both courses followed the depiction above. Each panel (representing a phase) was run as a separate batch of experiments. Within each batch, participants were randomly assigned across conditions within the phase. Participants were only allowed to participate in one phase (and hence one condition) per course. Arrows indicate dependencies between conditions across phases. Blue conditions did not require participants.
  }~\label{fig:experiment}
\end{figure}

Each of the two course began with a content phase in which 10 participants answered a set of questions for each lesson in the course and gave explanations for their answers. This phase simulates populating the content of a forum. This phase was followed by a voting phase that had 4 conditions with 25 participants in each condition.  In the voting phase, participants voted on which explanations they thought were good.  One condition in this phase had independent votes and three conditions had votes that were public to others present in that condition. Finally, there was a testing phase that had 6 conditions with 100 participants in each condition, with one condition for each of the voting phase conditions in addition to an expert graded condition and a random condition. Each phase is described in more detail below; see Figure~\ref{fig:experiment} for a flowchart depicting the different phases and conditions.

\begin{figure}[t]
\centering
  \includegraphics[width=\columnwidth]{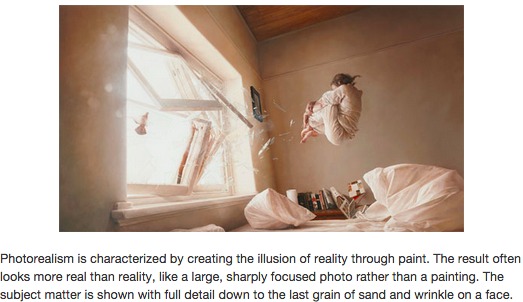}
  \caption{Depicts a sample lesson  from the Art course.
  Each lesson contained a brief description of two styles of art along with a sample painting corresponding to each, as depicted. The sample lesson for the Computer Science course contained a brief description of the topic along with a sample piece of pseudocode and its correct output.
  }~\label{fig:lesson}
\end{figure}

\begin{figure}[t]
\centering
  \includegraphics[width=\columnwidth]{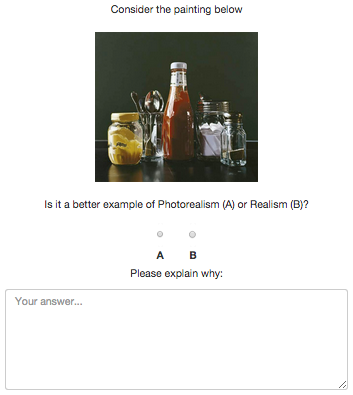}
  \caption{Depicts a sample question from the content phase of the Art course (for the lesson on Photorealism and Realism). The participant was shown a painting, asked whether it represented a painting of type A or B, and asked to give an explanation for their answer.}~\label{fig:exp1question}
\end{figure}

\begin{figure}[t]
\centering
  \includegraphics[width=\columnwidth]{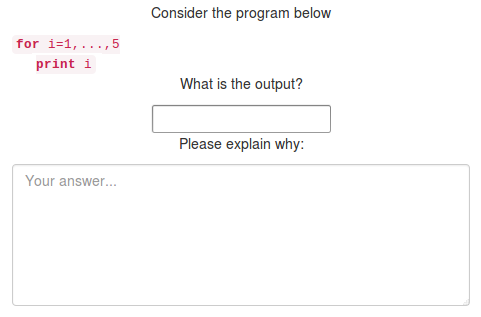}
  \caption{Depicts a sample question from the content phase of the Computer Science course (for the lesson on For Loops). The participant is shown a short piece of pseudocode, asked to type the output of the code, and asked to give an explanation for their answer.
  }~\label{fig:cs2question}
\end{figure}

\subsection{Participants}

Our experiment participants were recruited on Amazon Mechanical Turk (AMT), an online crowdsourcing platform in which {requesters} can post discrete, typically short, tasks with replicates called {HITs}. If an AMT {worker} finds and accepts a HIT, the worker gets paid upon successful completion. In our setting, each phase of each lesson corresponded to a task, and each task had many HITs falling into each of our different experimental conditions. We conducted our experiments over a two-week period in May of 2015.  All HITs within a single phase of each lesson were posted simultaneously and completed in under 12 hours.

Workers within a course participated in at most one phase, but were allowed to participate in both courses if they so desired. We placed weak selection criteria on the AMT workers, allowing workers with a minimum of 50 HITs completed and at least 90\% of completed HITs approved (i.e., marked as correct by the job requester). The workers were payed \$3.5, \$3.5 and \$2 per HIT for the content phase, voting phase and testing phase respectively, leading to an effective hourly wage of \$6.49 as measured by AMT.\footnote{We expect the hourly wage to actually be higher due to known factors such as skimming, in which workers accept and work on multiple HITs at once \cite{G2012}.} In total, 1211 unique workers participated in the experiment.

\subsection{Experiment Phases}

\subsubsection{Content Phase}
During the content phase, participants were presented with brief 3-5 sentence lessons (as illustrated in Figure~\ref{fig:lesson}). In each lesson, the participants were prompted to read the lesson, answer a question on the material, and provide an explanation as to why they chose their answers. In the Art course participants were shown a painting and asked to identify the art style of the painting given two choices (as illustrated in Figure~\ref{fig:exp1question}). In the Computer Science course, the participants were asked to give the output of a short piece of pseudocode (as illustrated in Figure~\ref{fig:cs2question}).

\begin{figure}[t]
\centering
  \includegraphics[width=.9\columnwidth]{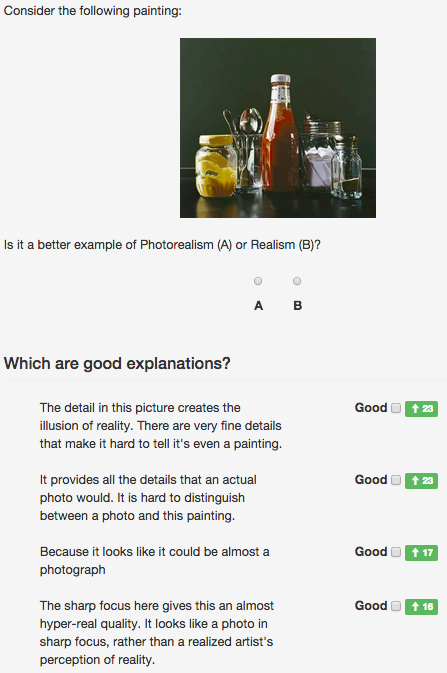}
  \caption{Depicts a sample voting phase  in the social condition. The independent condition does not display the green boxes containing the previous upvotes. Only a portion of the display is shown here. 10 explanations were presented at a given time, sorted according to the number of upvotes received thus far in the sequential condition and sorted randomly in the independent condition.}~\label{fig:exp2}
\end{figure}

\begin{figure}[t]
\centering
  \includegraphics[width=\columnwidth]{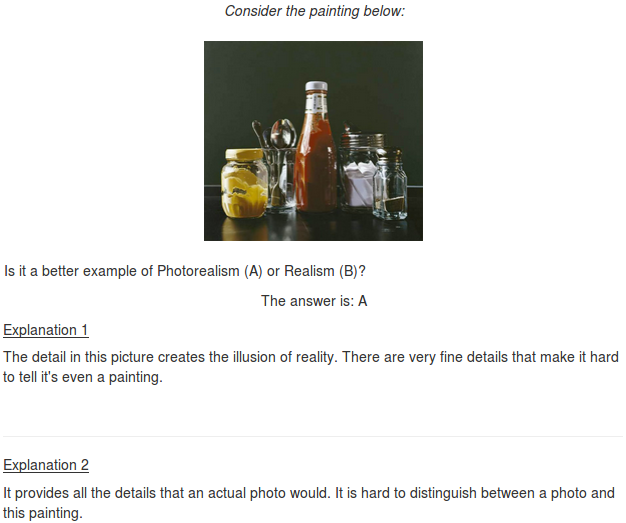}
  \caption{Depicts a sample testing phase: The student explanations, which were generated in the content phase and selected in the voting phase, are displayed in lieu of a lesson.}~\label{fig:exp3}
\end{figure}

\subsubsection{Voting Phase}
In the voting phase, participants were presented with the same brief lessons as in the content phase. The first part of the question was identical to the content phase, but instead of being asked to provide an explanation, participants were asked to upvote the explanations generated in the content phase that they thought were good (as illustrated in Figure~\ref{fig:exp2}). The format was modeled after common Q\&A platforms and MOOC forums.

This phase had two types of conditions, one in which the numbers of upvotes from previous participants were made public and the other in which these numbers were kept hidden. Public upvotes allow for {social influence} as participants can incorporate the information into their decisions to upvote.  This type of condition forms our {sequential condition}. We compared against a condition where the number of upvotes were private (same as Figure \ref{fig:exp2}, except the green boxes were not displayed), which we call the {independent condition}.

\subsubsection{Testing Phase}
During the testing phase, the goal was to measure a participant's ability to learn  {solely} from explanations generated in the content phase and chosen by the voting phase. No lesson was presented and vote information was not shown. Instead, the participants were presented with the questions asked in content and voting phases, along with two explanations from the content phase (as illustrated in Figure \ref{fig:exp3}).  The two explanations the participants were shown varied across conditions.

The participants were randomly assigned to one of four different types of conditions: sequential, independent, random and expert. The sequential conditions displayed the top two most upvoted explanations from the sequential conditions in the voting phase. We had three parallel replicates of the sequential condition. The independent condition displayed the top two most upvoted explanations from the independent condition of the voting phase. The random condition displayed two explanations selected uniformly at random from all the explanations generated in the content phase. In the expert condition, 
 the authors independently manually graded the content. The authors' grades were averaged into a final grade for each explanation, and the two explanations with the best average grades were presented. Ties were broken at random so that exactly two explanations were presented in every condition.

The tests given to participants consisted of four multiple-choice questions per lesson in the Art course similar to the one depicted in Figure~\ref{fig:exp1question}  but without the prompt for an explanation, and one numerical question per lesson in the Computer Science course. The computer science question consisted of a short piece of pseudocode and asked for its output, again without any explanation. 
As all questions had an objective right answer, each question was worth 0 or 1 point, depending on whether the answer was incorrect or correct.

\subsection{Self-Reported Learning}

Participants in all conditions were presented with the same optional survey at the end of the course, which included a self-assessment of learning. Participants reported their skill levels on the topic (Computer Science or Art) before and after the course on a scale from 0 (none) to 4 (expert). We measure self-reported learning by taking the difference between the before and after self-reported skill levels. While individual estimates of learning may be inflated overall, differences across conditions must be real since participants were randomly assigned to conditions.

\subsection{External Validators}

We further had external validators in art and computer science grade the student explanations in order to independently measure the quality of the content curated by the different conditions. This additional step was conducted due to the unexpected finding (discussed below) that the expert-curated condition did not outperform the other conditions, and was sometimes significantly worse. The external validation confirmed that the expert condition had selected good content as evaluated by external graders and lends weight to the finding. 
Moreover, the topics and question format were confirmed by the external validators to be similar to what is used in their courses.

The external validators were professors who had no previous knowledge of the experiment and no prior collaboration with the authors. Each has more than 15 years experience teaching introductory courses in their field, and either has developed or is in the process of developing a MOOC on such material. The external validators were asked to grade on the ``usefulness of the explanation to other students who will have to answer similar questions'', and  assigned a grade between 0-10 to each explanation generated in the content phase (100 explanations in Art, and 50 explanations in CS). 

\section{Related Work}

There are a number of avenues of research in computational social science related to our study.
Our work is related to the large literature on identifying content quality, which
primarily strives to develop automated techniques for quality
prediction in online settings. One way to identify the quality of content is based on the
contributor's reputation and past work \cite{CZW2006,DP2008}. Specific
content features can also be used, such as the inclusion of references
to external resources, the length, the utility or the verifiability
\cite{AZBA2008,K2007,KO2009}.  Another growing related area involves
popularity prediction \cite{HL2009,SH2010,cheng2014can}, which also
strives to quantify the effect of social influence on popularity
\cite{salganik2006experimental,KCPP2012,stoddard2015popularity}.
Other work has examined the impact of social influence on various
online user behaviors more generally
\cite{salganik2008experimental,C2008,muchnik2013social,van2014field,WWW2014}.
Our work contributes to these areas by identifying the effect that
social influence has on the discovery of high
quality content in a domain where quality can be objectively defined.

There are many mechanisms used for social recommendation systems. Variations, both on the allowable user input and on the form of
aggregation, exist. These variations have been studied widely in
theory and practice, and an overview is outside the scope of this work
(see, e.g., \cite{askalidis2013theoretical} and references cited
within). We select a specific form of of user input (upvotes on ranked
content) and the simplest and most common form of aggregation (sorting
by number of upvotes) since these choices are prevalent in
the education context. Most popular MOOC forums, including Coursera, MIT-X,
  Harvard-X, and Stanford Online, as well as {Q\&A} websites such as
  Yahoo! Answers, Stack Overflow, Baidu Knows and Quora  use
  variations of this type of sequential voting mechanism.

\section{Analysis}

We now present an overview of the participant responses followed by the results of our experiments using both quality metrics, self-reported learning and test scores.\footnote{The code and data used for our analysis are available online: {https://github.com/pkrafft/Sequential-Voting-Promotes-Collective-Discovery-in-Social-Recommendation-Systems}.} 

\subsection{Descriptive Statistics}

The conditions for the testing phase were the sequential conditions ($3$ replicates), the independent condition
($1$ replicate), the random condition ($1$ replicate) and the expert-curated condition ($1$ replicate). We used three runs of
the sequential condition because prior work suggests that social information may lead to different outcomes across replicates \cite{salganik2006experimental}.  These different outcomes occur because of herding amplifying the first few votes in each replicate.

100 HITs for each testing condition were posted on AMT; i.e., 600 HITs for the Art course and 600 HITs for the Computer Science course. The dataset was filtered so that only the results from workers who had completed all five lessons were retained. In some cases we had more participants than HITs, presumably from workers who clicked on our ad but did not accept the HIT. The number of completed lessons per condition ranged from 97 to 106 with an average of 100.92.

\begin{figure}[t!]
\centering
  \includegraphics[width=\columnwidth]{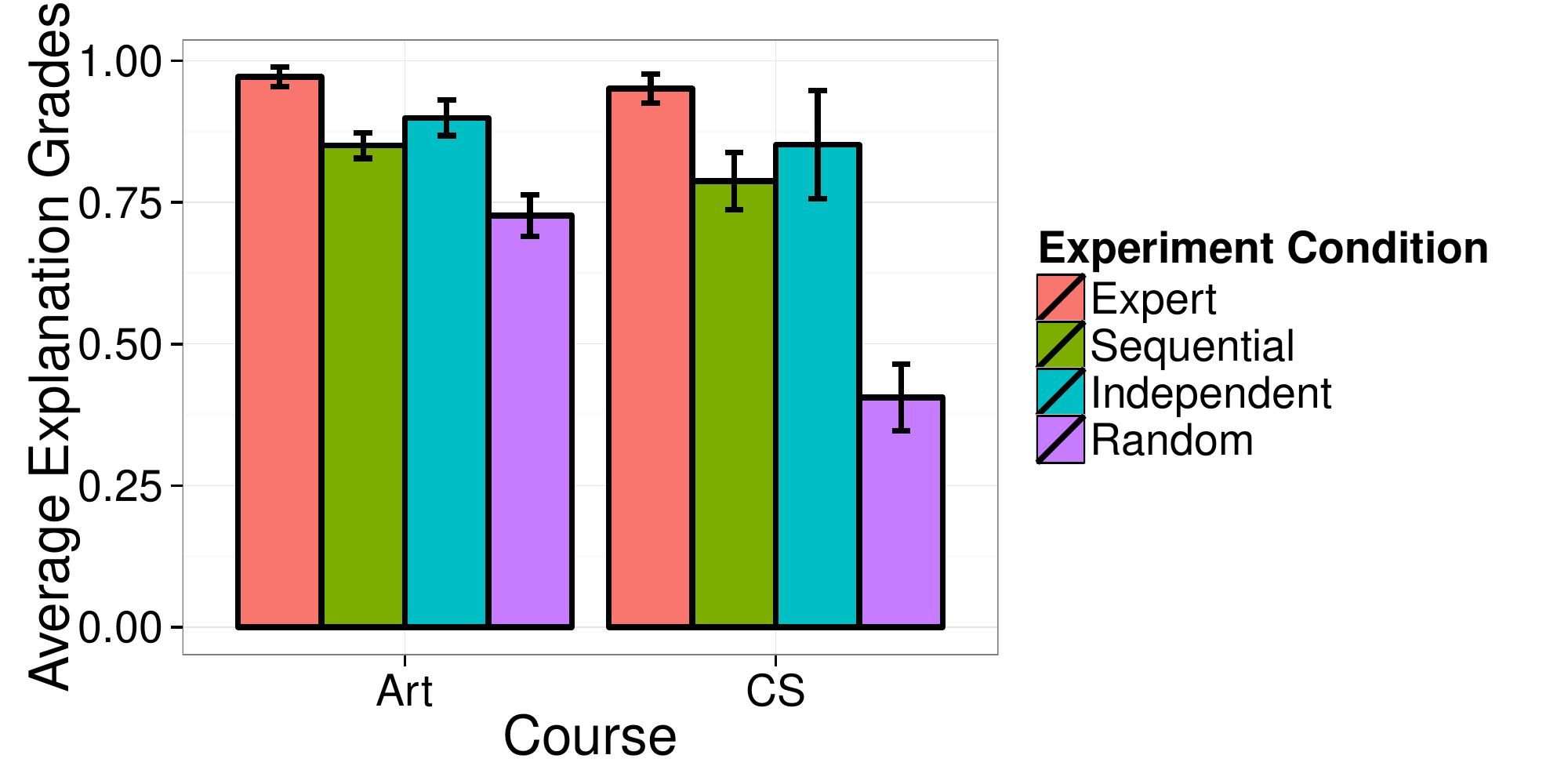}
  \caption{  Explanation grades given by external validators normalized per-lesson to a 0-1 scale. Error bars are standard errors of the mean. The plot shows that from the perspective of the external validators, explanations selected by the sequential and independent conditions are indistinguishable, while the explanations selected in the expert condition tend to be better and the random condition tends to be worse; despite this, we later see that the sequential condition tends to outperform the rest with respect to self-reported learning and demonstrated skill.
  }~\label{fig:grades}
\end{figure}

\begin{figure}[t!]
\centering
\includegraphics[width=0.625\columnwidth]{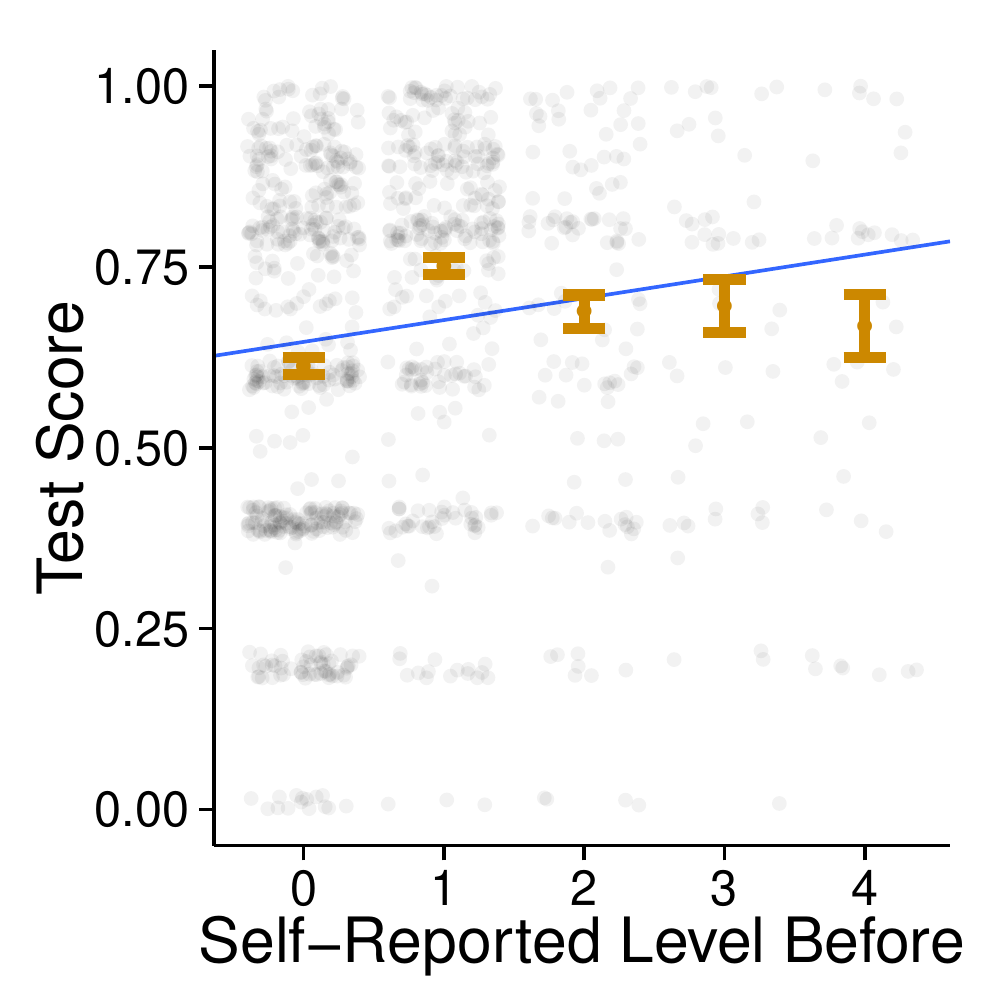}
\includegraphics[width=0.625\columnwidth]{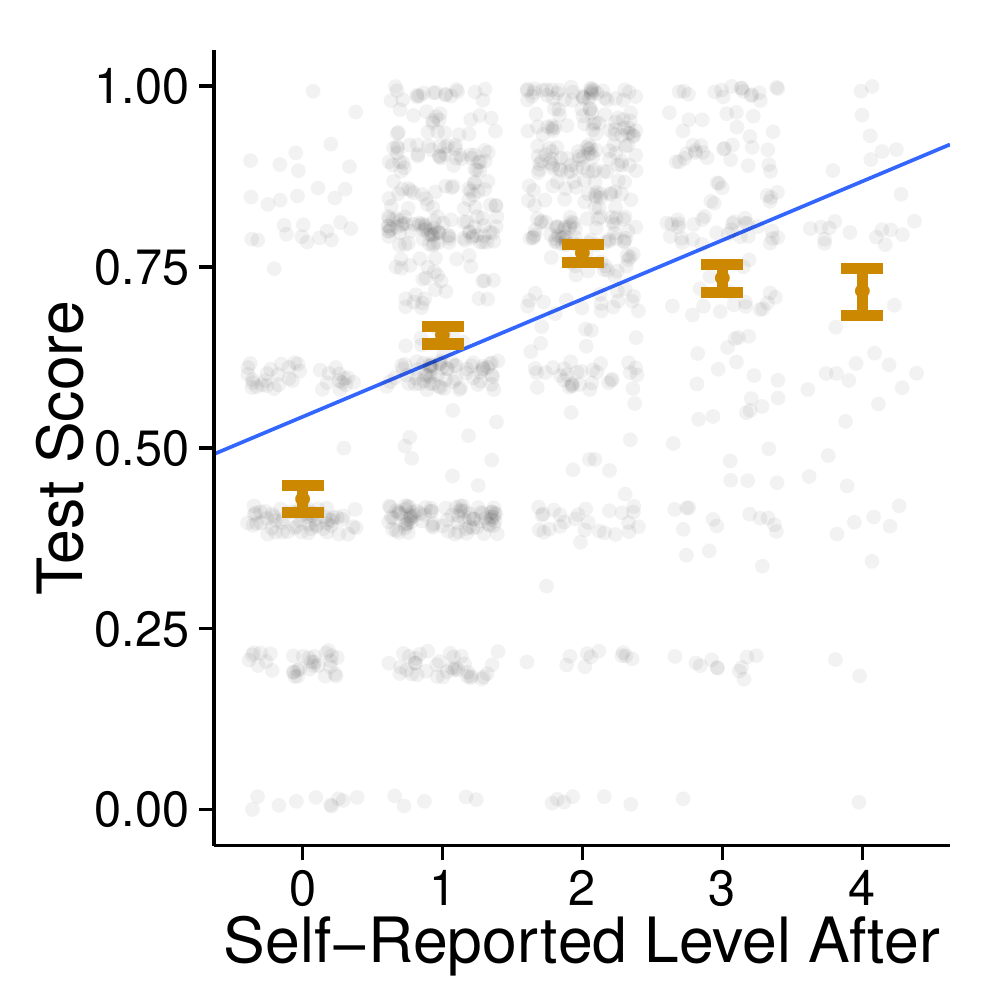}
  \caption{Self-reported skill levels before and after task completion, versus test scores. The test scores are more correlated with self-reported skill after a lesson, suggesting  that self-reported learning can be thought of as a proxy for actual learning. Error bars are standard errors of the means.  Regression lines are fitted to the raw data.}~\label{fig:levels}
\end{figure}

\subsection{Explanation Quality}

Our external validators independently graded all student explanations on a 0-10 scale. The combined score is an explanation's {raw grade}. If we think of the external validators as instructors in the course, this would be the grade they assign to an explanation posted on the course forum. 

The overall average explanation grade in the Art course was 6.13, and the average best explanation grade (across lessons) in this course was 8.45. The average explanation grade in the Computer Science course was approximately 2.36, and the average best explanation grade in the Computer Science course was 5.7.

We use these grades to evaluate the quality of the content selected by the different conditions.  We focus on the average grades of the two explanations selected in each condition of the voting phase. We normalize the grades so that, for each lesson, they lie on a 0-1 scale by dividing by the maximum grade in a given lesson.\footnote{This is necessary as some lessons are naturally harder than others. Without normalization the data across courses would not be comparable.} We then compared the average grade of the top explanations selected by each condition  (see Figure~\ref{fig:grades}).

In both the CS and the Art courses the explanation grades in the expert condition are significantly higher than the explanation grades in the sequential condition (two-sided $t$-test with $p =  0.009976$ for Computer Science and $p = 0.0001374$ for Art). 
We include this result to illustrate the differences in explanation quality as judged by an external validator. In particular, this shows that there is a range in explanation quality (especially in the Computer Science course).  This result also validates the authors' grades as used in the expert-curated condition. Though not statistically significantly, independent votes achieve identify higher graded explanations than sequential votes in both courses, as would be suggested by the literature on social influence. 
However, our desired metric was the effect on the {student}, i.e., the test scores and self-reported learning of a content consumer, not the grades given by experts.

\subsection{Learning}

In the testing phase we took explanations selected in the 
voting phase and assessed how well participants learned when consuming 
that material as instruction. We measure the change
in self-reported skill level from before and after the lessons. In addition, we measure the 
participants' knowledge with a test on the subject material after the lesson was completed.

The self-reported skill level before and after the course was submitted on a 0-4 scale with 0 being ``none'' and 4 being ``expert''. The test scores are normalized to be from 0 to 1. 
The self-reported skill level after task completion has a 
much higher correlation with test scores than the self-reported skill level 
from before the task (0.31186 and 0.11505 respectively; see Figure~\ref{fig:levels}). Hence, self-reported skill  after the course is a better proxy for actual skill level than self-reported skill level before the course.

\begin{figure}[t!]
\includegraphics[width=\columnwidth]{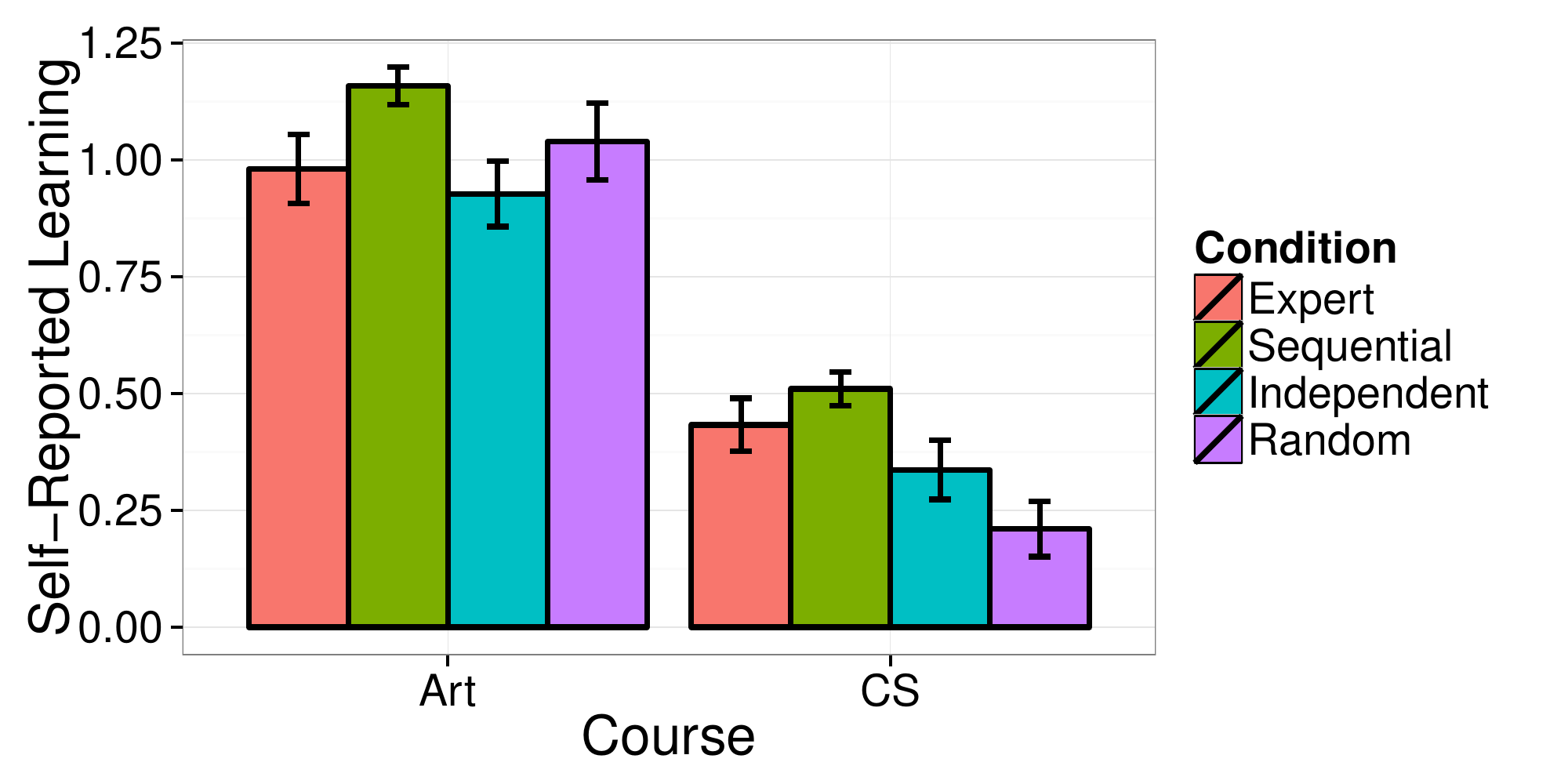}
  \caption{Comparison of self-reported learning scores.  Error bars are standard errors of the mean. The sequential condition  produces a significant increase in self-reported learning compared to the other conditions. 
  }~\label{fig:results}
\end{figure}

We find that self-reported learning in the sequential condition
is significantly higher than in the independent condition (two-sided $t$-test $p <
0.0005312$), in the random condition (two-sided $t$-test $p = 0.002109$), and in the expert
condition (two-sided $t$-test $p = 0.04359$). These differences remain significant
controlling for the course type and the experiment instance using
mixed effects models (with fixed effects for the course type and
random effects for the particular experiment instances). {Breaking the
analysis up by course, the trends are the same in all comparisons (see
Figure \ref{fig:results}), however the comparison to the expert
condition is not significant in the Computer Science course, and the comparison to
the random condition is not significant in the Art course.}

\begin{figure}
\hspace{.1in}  \includegraphics[width=\columnwidth]{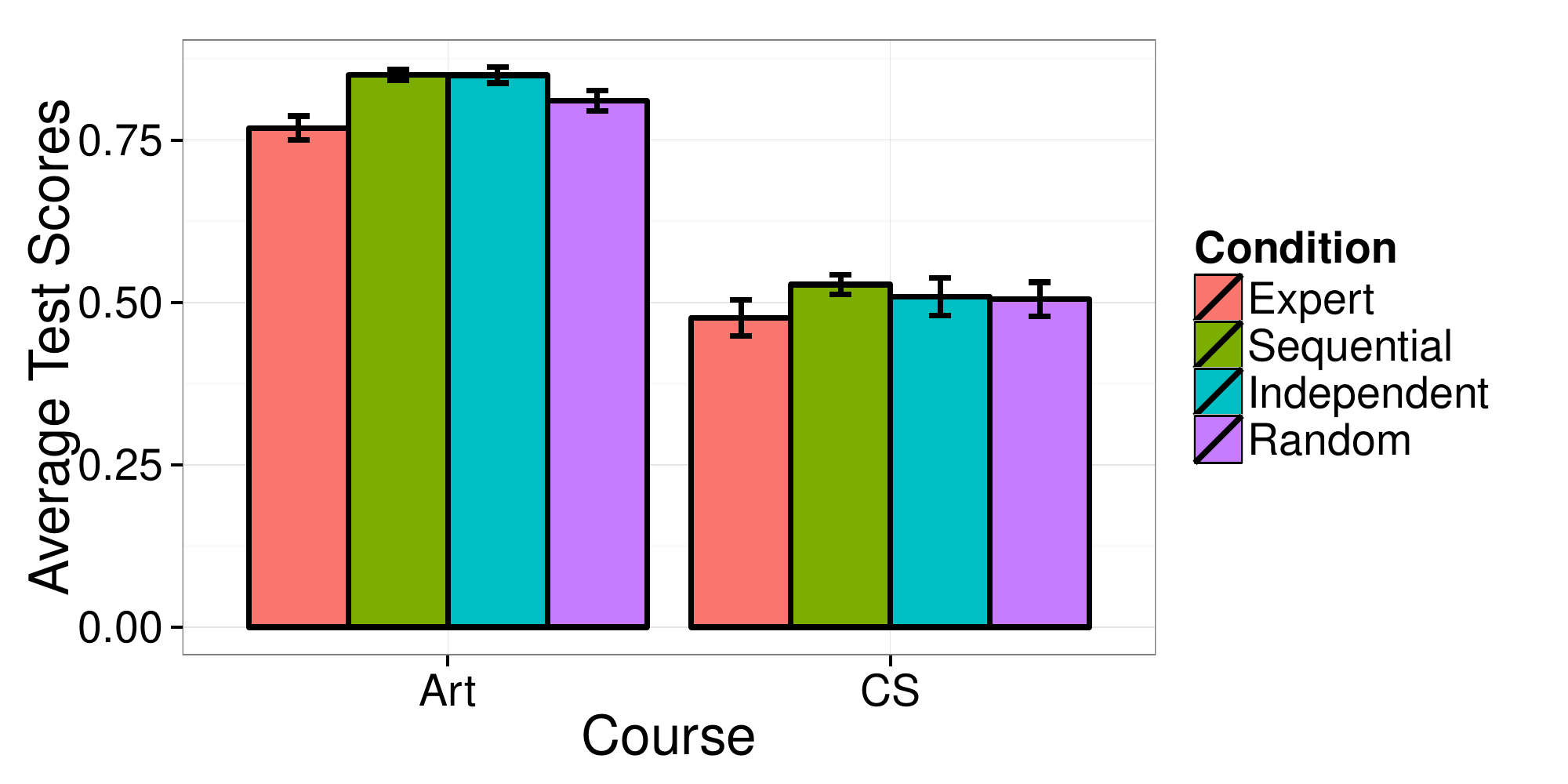}
  \caption{  Comparison of test scores.  Error bars are standard errors of the mean.  The sequential condition remains on average the best in terms of performance in both courses, though not always significantly. 
  }~\label{fig:results2}
\end{figure}

{Consistent with these trends, average test scores were higher in the
sequential condition than in the independent, random, and expert
conditions, though these comparisons were not statistically
significant (see Figure~\ref{fig:results2}).} The sequential condition is significantly
better than the expert condition (two-sided $t$-test $p = 0.02682$) and random (two-sided $t$-test $p < 
0.0001$) in the Art course, though not the Computer Science course.  Those single
comparisons are significant.  Using the mixed effects model to look at
both courses jointly, the comparison to the expert condition, but not to random,
remains significant.
The effect is weaker in one context (CS) than the other (Art) because the sequential and independent conditions were more similar in the CS test phase. Unlike in the Art course, the CS voting phase resulted in similar top explanations across sequential and independent conditions, likely because the content-generation phase resulted in fewer good explanations.

\subsection{Herding}

We hypothesized that the differences we observed in learning across conditions was due to a beneficial herding effect which allows for greater efficiency in identifying good explanations.  To test this hypothesis, we examined the skewness of the distribution of the ultimate number of votes each explanation got per question.  Following previous work \cite{gini1912,salganik2006experimental}, we first take the number of votes each explanation got and divide those numbers by the total number of votes per question.  These normalized popularity values are the ``market shares'' of popularity of each explanation.  We then use the average difference in market share between items as a measure of the skewness of the distribution, or the ``inequality'' in popularity.  Higher inequality values in the sequential condition compared to the independent condition indicates that herding may be occurring.

\begin{figure}
\centering
\includegraphics[width=.8\columnwidth]{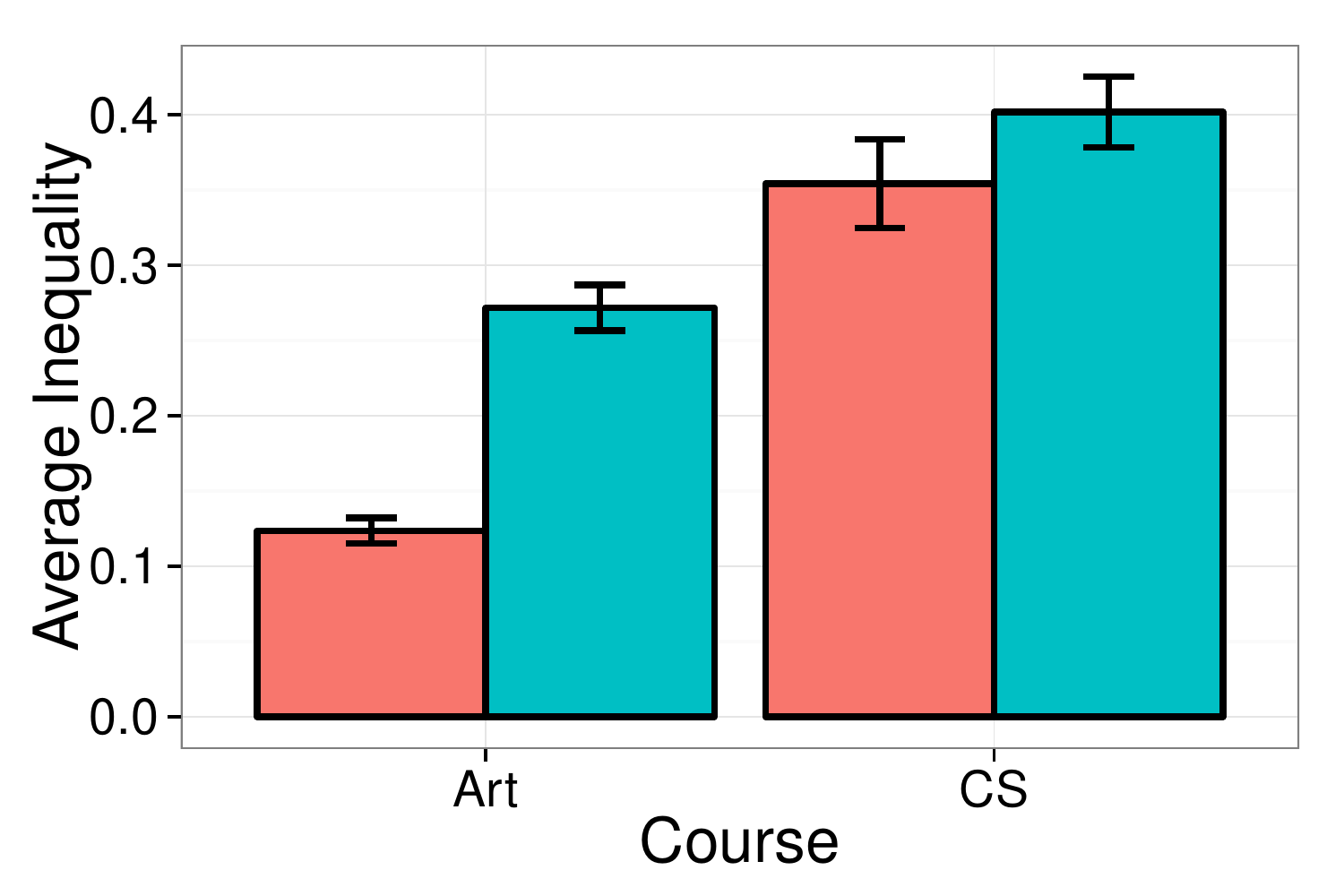}
  \caption{  Average inequality (Gini index) over the market share of upvotes. Error bars are standard errors of the mean. Higher inequality values in the sequential condition compared to the independent condition suggest that herding may be occurring.
  }~\label{fig:inequality}
\end{figure}

We find in the Art course that average inequality across questions is significantly higher in the sequential condition than the independent condition (two-sided t-test p $<$ 1e-5), suggesting a herding effect may be at play.  We also find a trend in the same direction in the Computer Science course, but the difference is not statistically significant (see Figure \ref{fig:inequality}).

\section{Discussion}

\paragraph{Main Results.}

The results using our objective measurements of quality paint a different picture than 
the grades of the external validators. 
Participants' self-reported assessments suggest that the sequential condition curates better content than all other conditions. 
This finding is notable since the sequential condition outperforms the independent condition (which had equivalent grades from the external validators), and since the sequential condition is not outperformed by the expert condition (which had significantly higher grades from the external validators). 
The differences in self-reported learning across conditions cannot be due to participants having a  tendency to tell us what we want to hear (i.e., report learning when none occurred) because of the randomization of participants to different  conditions.

Self-reported learning is an important metric since perceived progress is strongly correlated to motivation~\cite{S2003}. Moreover, self-reported learning can often be used as a valid proxy for more direct measures of learning~\cite{A1999}.  Therefore, observing an
improvement in self-reported learning alone is enough reason to adopt
the sequential voting procedure.

\paragraph{Limitations.}

{Participants' actual test scores follow the trend observed in the self-reported assessments, though not always with statistical significance.} While we did not expect to be under-powered, we believe that, due to the differences in experimental design between the two measurements, we were underpowered on test scores and not on self-reported learning. 
{Test scores used a standard between-subjects design while self-reported learning used a quasi-pretest-posttest between-subjects design (where the ``pretest'' and ``posttest'' were both end-survey questions).  The latter design should yield lower variance in the dependent variable, and thus will have higher power. A quasi-pretest-posttest (as opposed to a true pretest-posttest) was used so that participants would have the same context with which to address their skill. Having a (quasi)-pretest-posttest design was feasible for self-reported learning but not for actual test scores, since taking a test requires more time than completing a survey.}  
This is one of the main limitations of our work, and further investigation is required in order to determine whether the results on test scores are indeed  significant.

In our statistical tests, we used a $p$-value cutoff of 0.05 for declaring significance. If we instead use a Bonferroni correction, then the $p$-value cutoff would become 0.001.  This correction accounts for 50 tests, which is larger than the 30 tests in our final script for analyzing the data and larger than the 17 that we report. Using this substantially more conservative $p$-value, we still find that the sequential condition has significantly higher self-reported learning than the independent condition, but notably, the comparisons to the expert condition are no longer significant.

\paragraph{Domain Differences.}

As is readily apparent in Figures~\ref{fig:results} and \ref{fig:results2}, both the test scores and self-reported learning are lower in the Computer Science course than in the Art course. The disparity could be accounted for by the fact that the explanations in the Computer Science course were much worse than those in the Art course according to the grades assigned by the external validators (see the section on Explanation Quality); in other words, the content consumers had overall lower-quality explanations to help them learn, which could explain the lower learning scores overall.

\paragraph{Inadequacy of Experts.}

 In some cases, our evidence suggests that the sequential condition is better than the expert condition; in fact, in some cases even the random condition was on average (although not significantly) better than the expert condition. A priori, this result is very surprising, and initially we were concerned that one reason for it could be that the expert grading was performed poorly. The external validators were brought on at this point to grade the content in order to independently validate the experts' decisions. The fact that the external validation agreed with the author's decisions gives weight to the conclusion that sequential voting can in fact outperform experts; hence we do not believe this is a limitation of our work, but rather a point for further reflection. 
 
 One possible explanation for why expert-selected content could be worse is the mechanism used for curation: The experts did not vote on which {pairs} of explanations would be best.  They graded each explanation individually, and the explanations with the top two grades were selected for use in the expert condition. It is possible that sequential voting allows for better {joint} selection of explanations compared to a traditional grading method. 
%

\paragraph{Herding.}

We hypothesized that the differences in learning between the sequential and independent conditions was due to a beneficial herding effect that led to better ranked content.  
%
 Higher inequality values, measured with the Gini index \cite{gini1912}, in the sequential condition compared to independent conditions indicate that herding may be occurring.  However our findings were not conclusive in the CS course. The participants in the CS course only generated a few good explanations, while the participants in the Art course generated many. This factor likely led to high ``inequality'' in the distribution of votes for CS even in the independent condition, thereby diminishing the effect of herding.

\section{Conclusion \& Future Work}

Despite many models of collective intelligence clearly predicting
deficits in aggregate outcomes due to social influence, our work suggests that sequential voting
may indeed improve ranking and collective discovery in online social recommendation systems.
However, the mechanism through which sequential voting has 
this positive effect remains unclear.  While there is some evidence that herding may
be playing a role, for example, by emphasizing content that might have
been overlooked in the independent voting setting, another possibility is
that people are responding to others with anticorrelated votes in an intelligent manner; behavior that has recently been observed empirically \cite{SGJ2014}.  
Our understanding of collective intelligence, even in this simple interaction context, is clearly in a nascent stage; better models and further analyses are required.

Future work that uses the education domain to achieve objective
measurements of underlying quality should also address some of the
limitations of our present work.  For example, we see teasing out the
differences between self-reported learning and test scores as an
important topic for future work. More comprehensive tests, delays
between learning and testing, a true pre-test / post-test design, and
larger sample sizes of students could be employed in order to identify
whether our null results on test scores are actually a false negative.
Interesting high-level questions in this vein include: are participants
implicitly distinguishing between what they were {taught} as
opposed to what they {learned} in the self-assessment? How do
test scores and self-reported learning correlate, if at all, with
reported {confidence}? What type of model (if any) of social influence is warranted
by the results we observed, and can we increase the effects to produce better
outcomes? And what other forms of information can we make public in a
way that improves learning?

Our work also leaves open the mechanism through which information assimilation benefits learning. Understanding this mechanism would allow us to predict how student knowledge varies as a function of the type and amount of information revealed, and could lead to designing systems which maximize beneficial information, e.g., ranking content (as in \cite{lerman2014leveraging}) using learning as a quality metric.

The difference we observed between the expert and sequential conditions
also warrants further investigation.  This finding, if general,
suggests that sequential voting might lead not just to the crowd being
able to discover better content than it could through standard wisdom-of-crowds aggregation, but also better content than selected via expert grading.  Identifying the conditions under which crowds can outperform experts is an important open area of research in collective
intelligence, and our work suggests that even a simple interaction
mechanism may at times achieve this goal.



\section{Acknowledgments}

We would like to thank the anonymous reviewers for their insightful comments which have aided in the exposition of this work. 
We would also like to thank our three external validators: Prof. Nicolas Bock (University of Lausanne), Prof. Jean-C\'edric Chappelier (EPFL) and Prof. Djamila Sam-Haroud (EPFL), 
and the anonymous Amazon Mechanical Turk workers who participated in our study.  
The authors obtained approval for the experiment from the \'Ecole Polytechnique F\'ed\'ereal de Lausanne (EPFL) Human Research Ethics Committee (HREC) in Switzerland.
This material is based upon work supported by the National Science Foundation
Graduate Research Fellowship under Grant No. 1122374. Any opinion,
findings, and conclusions or recommendations expressed in this
material are those of the authors(s) and do not necessarily reflect
the views of the National Science Foundation

\bibliography{mini-mooc}
\bibliographystyle{aaai}

\end{document}